\documentclass[12pt,twocolumn]{iopart}

\expandafter\let\csname equation*\endcsname\relax

\expandafter\let\csname endequation*\endcsname\relax

\usepackage{amsmath}
\usepackage{graphicx}
\usepackage{dcolumn}
\usepackage{braket}

\def\bcen{\begin{center}}
\def\ecen{\end{center}}
\usepackage{epstopdf} 
\begin{document}

\title{Scheme for realizing quantum dense coding via entanglement swapping}

\author{Nilakantha Meher}
\address{Department of Physics, Indian Institute of Technology Kanpur, Kanpur 208016, India}
\ead{nilakantha.meher6@gmail.com}
\vspace{10pt}

\begin{abstract}

Quantum dense coding is a protocol for transmitting two classical bits of information from a sender (Alice) to a remote receiver (Bob) by sending only one quantum bit (qubit). In this article, we propose an experimentally feasible scheme to realize quantum dense coding \textit{via} entanglement swapping in a cavity array containing a certain number of two-level atoms. Proper choice of system parameters such as atom-cavity couplings and inter-cavity couplings allows perfect transfer of information.  A high fidelity transfer of information is shown to be possible by using recently achieved experimental values in the context of photonic crystal cavities and superconducting resonators. To mimic experimental imperfections, disorder in both the coupling strengths and resonance frequencies is considered. 
\end{abstract}

%
\vspace{2pc}
%
\submitto{\JPB}
%
\maketitle
%
%

\section{Introduction}
Quantum dense coding is an important task for realizing quantum communication \cite{Bennett}. It is a process of transmitting two classical bits of information from a sender (Alice) to a remote receiver (Bob) by sending only one qubit. This requires entanglement as an important resource. Due to potential application of quantum dense coding in quantum communication, a lot of attention has been paid to realizing this protocol in many physical systems such as  optical systems \cite{Mattle, Ban}, 
spin chains \cite{Liang, Simayi, Xu, Huang}, 
cavity quantum electrodynamics (QED) systems \cite{Xiu, Ye, Xue, Jia, Juan, Nie}, etc.

In the context of cavity QED, precise control of the system parameters makes the system as a promising candidate for realizing quantum information processing \cite{Reiserer}. Recently, a certain number of atoms dispersively interacting with a single cavity is shown to be a physical system capable of realizing quantum dense coding \cite{Ye,Juan, Xiu, Jia, Nie, Wu}. The importance of the dispersive coupling is to suppress the effect of dissipation.  In this configuration, Alice sends her atomic qubit to Bob for further processing to retrieve the information. On the other hand, Xue \textit{et. al.} proposed a scheme where the atom is taken as stationary qubit and the photon is used as a flying qubit in free space \cite{Xue}.

Due to recent progress in fabrication, it is possible to realize cavities with high quality ($Q$) factor. With the possibility of precise control of resonance frequencies and coupling strengths of the array, coupled cavities provide a suitable physical system for realizing many quantum information protocols \cite{Meher, Almeida,YangLiu,Yung,Meher_2019,Neto,Hua,Meher2019}. The scalability of the array is an additional advantage for realizing distance communication. Cavities have been used to realize many interesting phenomena such as entanglement generation \cite{Leons,Liew,Browne,Miry}, nonclassical state preparation \cite{Yurk,Rojan,WeiWei,Yanhua, Meher3}, localization-delocalization \cite{Meher2,Schmidt}, heat transfer \cite{Asad,Meher_2020,Xuereb}, photon blockade \cite{Imamoglu,Birnbaum,Tang,Shen,Miranowicz} etc by properly tuning the resonance frequencies or including material medium inside it. In addition, suitable choices of coupling strengths between the cavities provide perfect transfer of photons between the cavities \cite{Meher,Almeida,Zhou,Qin}.


In this article, we proposed a scheme to realize quantum dense coding protocol in an atom-cavity system. The system consists of a cavity array, and each end cavity of the array contains a two-level atom which is accessible by Alice and Bob respectively. Bob has an additional atom which is entangled with the atom of Alice. Hence, Alice has only one atom whereas Bob has two atoms. In order to transfer the information,  Alice encodes the information on her qubit (atom) by applying a unitary transformation. Then, Alice allows her qubit to interact with the cavity array. If the cavity-cavity and the atom-cavity coupling strengths are properly chosen, the entanglement between Alice's and Bob's qubits transferred to both the qubits of Bob. At this point, Alice's and Bob's qubits are disentangled where as Bob's qubits are entangled. This is referred to as entanglement swapping in the literature \cite{Yang, Ying,Xiu2007}.  Here the cavity array acts as a quantum channel and photon as the information carrier. Hence, the time scale for information transfer  depends on the inter-cavity coupling strengths. We find that the fidelity of the transfer of two bits of classical information from Alice to Bob is unity in the absence of dissipation. The master equation is employed to calculate the fidelity in the presence of dissipation. Using the experimentally achievable values in the context of photonic crystal cavities and superconducting resonators, we show that the fidelity of information transfer is high. To mimic experimental imperfection, disorder is included in the system.

This article is organized as follows: In Sec. \ref{PS}, we describe our physical system and provide the choices of coupling strengths for the perfect transfer of information.  Quantum dense coding \textit{via} coupled cavity array is presented in Sec. \ref{QDCC}. We discuss the results obtained by using experimentally achievable values of the system parameters in Sec. \ref{Results}. Finally, we summarized our results in Sec. \ref{Summary}.   
\section{Physical system}\label{PS}
\begin{figure*}
\begin{center}
\includegraphics[scale=0.48]{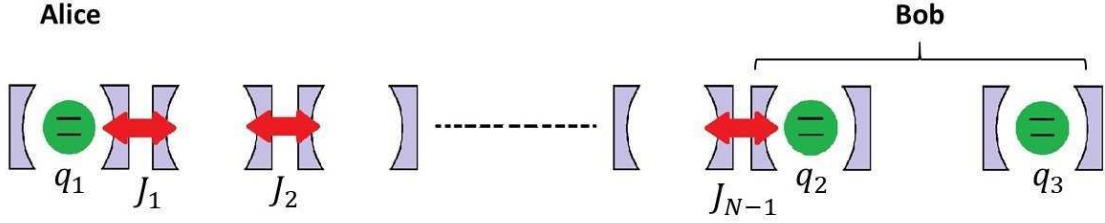}
\caption{Schematic of coupled cavity array whose end cavities contain one atom in each. Atoms $q_1$ and $q_2$ are accessible by Alice and Bob respectively. Bob has an additional cavity with an atom $q_3$, which are not connected to each other as well as not coupled to the array.  }
\label{PhysicalSystem}
\end{center}
\end{figure*}

We consider an array of $N$ coupled cavities as shown in Fig. \ref{PhysicalSystem}. The end cavities of the array contain two two-level atoms $q_1$ and $q_2$. Alice can control qubit $q_1$ which is present in the first cavity and Bob can control $q_2$ which is present in the $N$th cavity. In addition, Bob has another qubit $(q_3)$ in $(N+1)$th cavity, which is separated from the array. We assume that qubit $q_3$ does not interact with the $(N+1)$th cavity. Then the Hamiltonian for this system is
\begin{align}\label{Hamiltonian}
H &=\sum_{i=1}^3 \omega_{qi} (\sigma_+\sigma_-)_{qi}+\omega\sum_{k=1}^{N+1} a_k^\dagger a_k  \nonumber\\
&+\sum_{k=1}^{N-1} J_{k}(a_k^\dagger a_{k+1}+a_k a_{k+1}^\dagger)+g [a_1^\dagger (\sigma_-)_{q1}+a_1 (\sigma_+)_{q1}] \nonumber\\
&+g [a_N^\dagger (\sigma_-)_{q2}+a_N (\sigma_+)_{q2}],
\end{align}
where $\omega_{qi}$ is the atomic resonance frequency of the $i$th qubit. All the cavities are considered to have the same resonance frequencies $\omega$. We also assume $\omega_{q1}=\omega_{q2}=\omega$, \textit{i.e.,} both the qubits $q_1 $ and $q_2$ are resonantly interacting with their respective cavities. The coupling strength between the $k$th and $(k+1)$th cavities is $J_k$, and $g$ is the coupling strength between the qubits and their respective cavities. We choose the form of the coupling strengths $g$ and $J_k$ as
\begin{align}\label{EngineeredCoupling}
g=\sqrt{N+1}J,\nonumber\\
J_k=\sqrt{(k+1)(N+1-k)}J,
\end{align}
where $N$ is the number of cavities in the array and $J$ is a constant. Similar choices of coupling strengths are used for the perfect transfer of photon in a coupled cavity array \cite{Meher, Yogesh}. The operator $a_k (a_k^\dagger)$ is the annihilation (creation) operator for the $k$th cavity. The raising and lowering operators for the $i$th qubit are $(\sigma_+)_{qi}=(\ket{e}\bra{g})_{qi}$ and $(\sigma_-)_{qi}=(\ket{g}\bra{e})_{qi}$ respectively. Here, $\ket{g}_i (\ket{e}_i)$ is the ground (excited) state of the $i$th qubit.

A basic requirement for realizing quantum dense coding in our system is perfect transfer of a quanta between the qubits $q_1$ and $q_2$. If the qubit $q_1$ is in the excited state, and rest of the cavities and qubits are in their respective ground state, then the evolved state under the Hamiltonian $H$ is (for derivation refer the Appendix. A)
\begin{align}\label{timeevolutionSingeQuantum}
&e^{-iHt}\ket{e}_1\ket{\text{vac}}\ket{g}_2\ket{g}_3=e^{-i\omega t}\nonumber\\
\times &\left[(\cos Jt)^{N+1}\ket{e}_1\ket{\text{vac}}\ket{g}_2\ket{g}_3 \right. \left.+ \sum_{k=1}^{N} C_k\ket{g}_1\ket{k}\rangle\ket{g}_2\ket{g}_3 \right. \nonumber\\
&\left.+(-i\sin Jt)^{N+1}\ket{g}_1\ket{\text{vac}}\ket{e}_2\ket{g}_3\right],
\end{align}
where
\begin{align}\label{Ck}
C_k=\sqrt{\frac{(N+1)!}{(N+1-k)!k!}} (\cos Jt)^{N+1-k} (-i\sin Jt)^k .
\end{align}
Here $\ket{\text{vac}}$ is the state that represents all the cavities are in vacuum. The state $\ket{k}\rangle$ represents the $k$th cavity is in a single photon state and the rest of the cavities are in vacuum.
The probability of transferring a quanta from $q_1$ to $q_2$ is
\begin{align}
P=(\sin Jt)^{2(N+1)}.
\end{align}
The probability $P$ becomes unity at time $t=\pi/2J$, which we denote as $T$. This is the minimum time required to transfer a quanta between qubits $q_1$ and $q_2$.

\section{Quantum dense coding in the array}\label{QDCC}
In this section, we describe the dense coding protocol in our physical model.  We consider the initial state of the system as
\begin{align}\label{initialstate}
\ket{\psi(0)}=\frac{1}{\sqrt{2}}[\ket{g}_1\ket{\text{vac}}\ket{g}_2\ket{g}_3+\ket{e}_1\ket{\text{vac}}\ket{g}_2\ket{e}_3],
\end{align}
where the qubits $q_1$ and $q_3$ are  entangled. This state can be prepared experimentally \textit{via} entanglement swapping \cite{Yang, Ying,Xiu2007}. In order to encode the two classical bits in a single qubit, Alice applies unitary operation on her qubit depending on the choice of the classical bits. Table \ref{Operations}  shows the choices of classical bits and the corresponding unitary operations to be applied on Alice's qubit.
\begin{table}[h!]
\caption{Classical bits and unitary operations.}
\begin{center}
\begin{tabular}{ |c|c| } 
 \hline
 $(x,y)$ & unitary operation on $q_1$ \\ 
 \hline
 (0,0) & $I$\\ 
 \hline
 (1,0) & $\sigma_z$ \\ 
  \hline
 (0,1) & $\sigma_x$\\ 
 \hline
 (1,1) & $\sigma_z$ and $\sigma_x$ \\ 
 \hline
\end{tabular}
\label{Operations}
\end{center}
\end{table}

The actions of unitary operators on the qubit $q_1$ are $\sigma_z\ket{e}_1=\ket{e}_1$, $\sigma_z\ket{g}_1=-\ket{g}_1$, $\sigma_x\ket{e}_1=\ket{g}_1$, $\sigma_x\ket{g}_1=\ket{e}_1$. Then, the initial state given in Eqn. \ref{initialstate} after the unitary operation becomes
\begin{subequations}
\begin{align}\label{Operation1}
\ket{\psi_1}=\frac{1}{\sqrt{2}}[\ket{g}_1\ket{\text{vac}}\ket{g}_2\ket{g}_3+\ket{e}_1\ket{\text{vac}}\ket{g}_2\ket{e}_3],
\end{align}
\begin{align}\label{Operation2}
\ket{\psi_2}=\frac{1}{\sqrt{2}}[\ket{e}_1\ket{\text{vac}}\ket{g}_2\ket{e}_3-\ket{g}_1\ket{\text{vac}}\ket{g}_2\ket{g}_3],
\end{align}
\begin{align}\label{Operation3}
\ket{\psi_3}=\frac{1}{\sqrt{2}}[\ket{e}_1\ket{\text{vac}}\ket{g}_2\ket{g}_3+\ket{g}_1\ket{\text{vac}}\ket{g}_2\ket{e}_3],
\end{align}
\begin{align}\label{Operation4}
\ket{\psi_4}=\frac{1}{\sqrt{2}}[\ket{g}_1\ket{\text{vac}}\ket{g}_2\ket{e}_3-\ket{e}_1\ket{\text{vac}}\ket{g}_2\ket{g}_3],
\end{align}
\end{subequations}
where $\ket{\psi_1}=I\ket{\psi(0)}$, $\ket{\psi_2}=\sigma_z\ket{\psi(0)}$, $\ket{\psi_3}=\sigma_x\ket{\psi(0)}$, $\ket{\psi_4}=\sigma_x\sigma_z\ket{\psi(0)}$. Now, Alice allows the qubit $q_1$ to interact with the cavity array. Then, the above states evolve under the Hamiltonian $H$ as
\begin{subequations}
\begin{align}\label{EvolvedState1}
&\ket{\psi_1(t)}=e^{-iHt}\ket{\psi_1}=\frac{1}{\sqrt{2}}\left[\ket{g}_1\ket{\text{vac}}\ket{g}_2\ket{g}_3 \right. \nonumber\\
&+e^{-i\omega_{q3}t}e^{-i\omega t}\left[(\cos Jt)^{N+1}\ket{e}_1\ket{\text{vac}}\ket{g}_2\ket{e}_3 \right. \nonumber\\
&\left.+ \sum_{k=1}^{N} C_k\ket{g}_1\ket{k}\rangle\ket{g}_2\ket{e}_3 \right. \nonumber\\
&\left.\left.+(-i\sin Jt)^{N+1}\ket{g}_1\ket{\text{vac}}\ket{e}_2\ket{e}_3\right]\right],
\end{align}
\begin{align}\label{EvolvedState2}
&\ket{\psi_2(t)}=e^{-iHt}\ket{\psi_2}=\frac{1}{\sqrt{2}}\left[-\ket{g}_1\ket{\text{vac}}\ket{g}_2\ket{g}_3 \right. \nonumber\\
&+e^{-i\omega_{q3}t}e^{-i\omega t}\left[(\cos Jt)^{N+1}\ket{e}_1\ket{\text{vac}}\ket{g}_2\ket{e}_3 \right. \nonumber\\
&\left.+ \sum_{k=1}^{N} C_k\ket{g}_1\ket{k}\rangle\ket{g}_2\ket{e}_3 \right. \nonumber\\
&\left.\left.+(-i\sin Jt)^{N+1}\ket{g}_1\ket{\text{vac}}\ket{e}_2\ket{e}_3\right]\right],
\end{align}
\begin{align}\label{EvolvedState3}
&\ket{\psi_3(t)}=e^{-iHt}\ket{\psi_3}=\frac{1}{\sqrt{2}}\left[\ket{g}_1\ket{\text{vac}}\ket{g}_2\ket{e}_3 e^{-i\omega_{q3}t}\right.\nonumber\\
&+e^{-i\omega t}\left[(\cos Jt)^{N+1}\ket{e}_1\ket{\text{vac}}\ket{g}_2\ket{g}_3 \right. \nonumber\\
&\left.+ \sum_{k=1}^{N} C_k\ket{g}_1\ket{k}\rangle\ket{g}_2\ket{g}_3 \right. \nonumber\\
&\left.\left.+(-i\sin Jt)^{N+1}\ket{g}_1\ket{\text{vac}}\ket{e}_2\ket{g}_3\right]\right],
\end{align}
\begin{align}\label{EvolvedState4}
&\ket{\psi_4(t)}=e^{-iHt}\ket{\psi_4}=\frac{1}{\sqrt{2}}\left[\ket{g}_1\ket{\text{vac}}\ket{g}_2\ket{e}_3 e^{-i\omega_{q3}t}\right.\nonumber\\
&-e^{-i\omega t}\left[(\cos Jt)^{N+1}\ket{e}_1\ket{\text{vac}}\ket{g}_2\ket{g}_3 \right. \nonumber\\
&\left.+ \sum_{k=1}^{N} C_k\ket{g}_1\ket{k}\rangle\ket{g}_2\ket{g}_3 \right. \nonumber\\
&\left.\left.+(-i\sin Jt)^{N+1}\ket{g}_1\ket{\text{vac}}\ket{e}_2\ket{g}_3\right]\right].
\end{align}
\end{subequations}
 The expression for $C_k$ is given in Eqn. \ref{Ck}. We set $\omega/J=4n-(N+1)$ and $\omega_{q3}/J=4m$ ($n$ and $m$ are two integers) for subsequent discussion. Then  the evolved states at the time $T=\pi/2J$ are
 \begin{subequations}\label{EvolvedStatepi2J}
\begin{align}\label{EvolvedStatepi2J1}
\ket{\psi_1(T)}=\frac{1}{\sqrt{2}}[\ket{g}_1\ket{\text{vac}}\ket{g}_2\ket{g}_3+\ket{g}_1\ket{\text{vac}}\ket{e}_2\ket{e}_3], 
\end{align}
\begin{align}\label{EvolvedStatepi2J2}
\ket{\psi_2(T)}=\frac{1}{\sqrt{2}}[\ket{g}_1\ket{\text{vac}}\ket{e}_2\ket{e}_3-\ket{g}_1\ket{\text{vac}}\ket{g}_2\ket{g}_3], 
\end{align}
\begin{align}\label{EvolvedStatepi2J3}
\ket{\psi_3(T)}=\frac{1}{\sqrt{2}}[\ket{g}_1\ket{\text{vac}}\ket{e}_2\ket{g}_3+\ket{g}_1\ket{\text{vac}}\ket{g}_2\ket{e}_3], 
\end{align}
\begin{align}\label{EvolvedStatepi2J4}
\ket{\psi_4(T)}=\frac{1}{\sqrt{2}}[\ket{g}_1\ket{\text{vac}}\ket{g}_2\ket{e}_3-\ket{g}_1\ket{\text{vac}}\ket{e}_2\ket{g}_3]. 
\end{align}
\end{subequations}
It is to be noted that the qubits $q_2$ and $q_3$ become entangled in all these cases. In other words, entanglement between the qubits $q_1$ and $q_3$ was swapped to $q_2$ and $q_3$. The time $T=\pi/2J$ is the minimum time to transfer the information (swap the entanglement) and at this time Bob has to  decouple qubit $q_2$ from the array. To decode the information, Bob applies a CNOT operation on his qubits $q_2$ and $q_3$. The CNOT operation \cite{Divincenzo2,Divincenzo3, Barenco2} transforms the basis states as $U_{CNOT}\ket{g}_2\ket{g}_3=\ket{g}_2\ket{g}_3$, $U_{CNOT}\ket{g}_2\ket{e}_3=\ket{g}_2\ket{e}_3$, $U_{CNOT}\ket{e}_2\ket{g}_3=\ket{e}_2\ket{e}_3$ and $U_{CNOT}\ket{e}_2\ket{e}_3=\ket{e}_2\ket{g}_3$. The CNOT operation essentially disentangles qubits $q_2$ and $q_3$ \cite{Barenco2}. After the CNOT operation, Bob applies a Hadamard operation on qubit $q_2$. The Hadamard operation \cite{DiVincenzo} transforms the states as $U_H(\ket{g}_2+\ket{e}_2)/\sqrt{2}=\ket{g}_2$ and  $U_H(\ket{g}_2-\ket{e}_2)/\sqrt{2}=\ket{e}_2$. After the CNOT and Hadamard operations, the states given in Eqns. (\ref{EvolvedStatepi2J1}-\ref{EvolvedStatepi2J4}) become $\ket{g}_2\ket{g}_3$, $\ket{e}_2\ket{g}_3$, $\ket{g}_2\ket{e}_3$ and $\ket{e}_2\ket{e}_3$ respectively. Now, by mapping the states $\ket{g}\rightarrow 0$ and $\ket{e}\rightarrow 1$, Bob recovers the classical bits.\\

\begin{figure}
\begin{center}
\includegraphics[width=13cm, height=8cm]{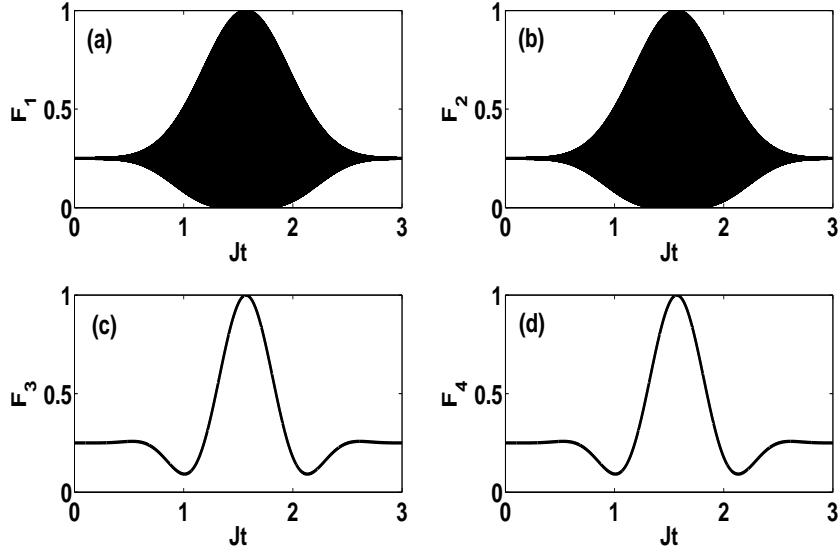}
\caption{Fidelities $F_1,F_2,F_3$ and $F_4$ as a function of $Jt$. We choose $\omega/J=9.995\times 10^3$, $\omega_{q3}/J=10^4$ and $N=4$. All the coupling strengths satisfy the relation given in Eqn. \ref{EngineeredCoupling}.}
\label{FidelityIdeal}
\end{center}
\end{figure}

It is to be noted that the probability of transferring the information depends on the evolution of the states from Eqns. (\ref{Operation1}-\ref{Operation4}) to Eqns. (\ref{EvolvedStatepi2J1}-\ref{EvolvedStatepi2J4}). Hence, the fidelity of information transfer is
\begin{align}
F_i=|\bra{\psi_i(T)}e^{-iHt}\ket{\psi_i}|^2. ~~~(i=1,2,3,4)
\end{align}
Fidelities are shown in Fig. \ref{FidelityIdeal}. It is seen that fidelities are unity at $T=\pi/2J$. We have chosen $\omega/J=4n-(N+1)$ and $\omega_{q3}/J=4m$ with $n=2500,m=2500$ and $N=4$. Hence, the choice of coupling strengths given in Eqn. \ref{EngineeredCoupling} ensures perfect transfer of information. It is seen that $F_1=F_2$ and $F_3=F_4$. 
\section{Results and discussion}\label{Results}
In this section, we will discuss the quantum dense coding protocol by considering experimentally achievable parameters. As dissipation is unavoidable, we include the dissipation in the system. Loss of photons may occur via spontaneous emission or leakage through the cavity walls. The effect of dissipation on the fidelity of information transfer is studied by analyzing the master equation. The master equation that includes the dissipation in the system is \cite{Carmichael}
\begin{align}
\frac{\partial \rho}{\partial t}=\frac{-i}{\hbar}[H,\rho]+\frac{k}{2}\sum_{j=1}^{N+1}\mathcal{L}[a_j]\rho+\frac{\gamma}{2}\sum_{j=1}^3\mathcal{L}[(\sigma_-)_{qj}]\rho,
\end{align}
where $\mathcal{L}[o]\rho=2o \rho o^\dagger-o^\dagger o\rho-\rho o^\dagger o$ is the Lindblad superoperator \cite{Lindblad}, and $k$ and $\gamma$ are the dissipation rates of cavities and atoms respectively. In order to minimize the effect of dissipation on fidelities, a cavity array with high quality factors ($Q$-factors) is necessary. High $Q$ value implies that the photon life time inside the cavity is large. Recently, photonic crystal cavities with large $Q$ factor have been fabricated and the order of $Q$-factor was $\sim 10^6$ \cite{Noto,Noto2,Vasco}. Another factor that can minimize the effect of dissipation is the coupling strength. If the cavities are highly coupled then the time for the information transfer will be small, as a result the effect of dissipation will be small. Highly coupled cavities are already been realized in the context of photonic crystal cavities whose coupling strengths are of the order of $\sim$ THz \cite{Arka}. Hence, an array of photonic crystal cavities may be a suitable physical system for realizing this protocol. Another platform which is suitable for realizing the quantum dense coding protocol is a superconducting resonator array. It is possible to fabricate superconducting resonators with $Q$-factors larger than $10^6$ \cite{Kuhr}. 
\begin{figure}
\begin{center}
\includegraphics[scale=0.30]{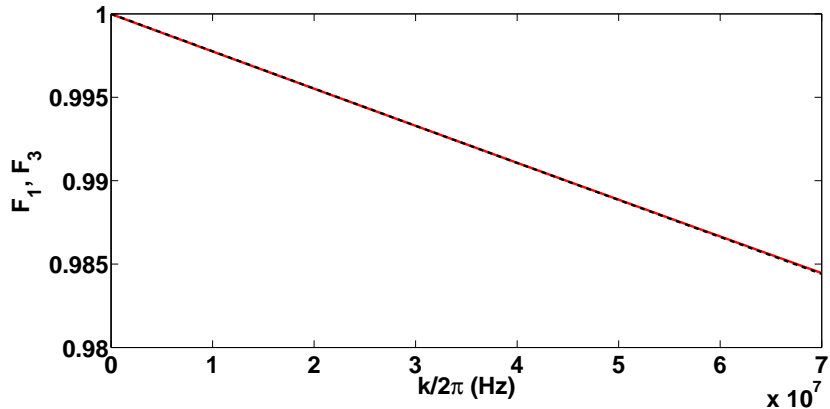}
\caption{Fidelities $F_1$ (continuous) and $F_3$ (dashed) at time $T=\pi/2J$ are shown as a function of cavity decay rate $k/2\pi$. We choose $(\omega/2\pi,\omega_{q3}/2\pi)=(6.92, 7)$THz, $(J/2\pi, g/2\pi)=(7,23.21)$GHz, $\gamma/2\pi=3.5$ MHz and $N=10$.}
\label{DecayFidelity}
\end{center}
\end{figure}

For the purpose of demonstration of quantum dense coding, we plot $F_1$ and $F_3$ as a function of $k/2\pi$ in Fig. \ref{DecayFidelity} by setting $(\omega/2\pi, \omega_{q3}/2\pi)=(6.92,7)$THz, $(J/2\pi,g/2\pi)=(7,23.21)$GHz and the atomic decay rate $\gamma/2\pi=3.5$ MHz. These values are easily achievable in photonic crystal cavities \cite{Arka,Arka2,Xiao,Tiecke,Du}. As seen from the figure, the fidelity of information transfer is more than $0.98$ even for large dissipation rate. We have taken $N=10$. The time for information transfer is $\pi/2J\sim 0.22$ns, which is much smaller than the cavity decay time $2\pi/k\sim 14$ns for $k/2\pi=70$ MHz. In the case of superconducting resonators, we set $J/2\pi=1.9$ MHz, $\omega/2\pi=1.88$ GHz, $\omega_{q3}=1.9$ GHz and the coupling between the resonator and superconducting qubit is taken to be $g/2\pi=6.3$ MHz \cite{Ma, Chen, Li2}. The decay parameters are chosen to be $k/2\pi=1.8$ kHz and $\gamma/2\pi=1$ kHz. The number of resonators in the array is $N=10$. Using these values, the fidelity of information transfer is found to be $0.992$. In this case, the time for information transfer is $0.82\mu$s while the resonator relaxation time is $0.87$ms.
\begin{figure}
\begin{center}
\includegraphics[scale=0.30]{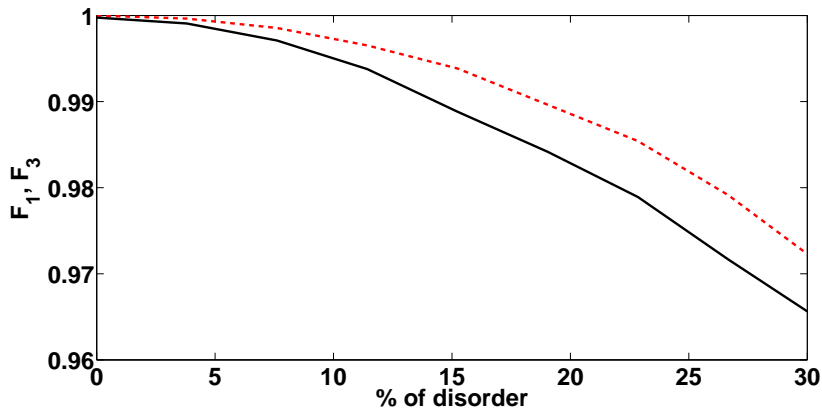}
\caption{Fidelities $F_1$ (continuous) and $F_3$ (dashed) at time $T=\pi/2J$ are shown as a function of percentage of disorder. We choose $(\omega/2\pi,\omega_{q3}/2\pi)=(6.92, 7)$THz, $(J/2\pi, g/2\pi)=(7,23.21)$GHz and $N=10$.}
\label{FluctuationVsFidelity}
\end{center}
\end{figure}
\begin{figure}
\begin{center}
\includegraphics[scale=0.30]{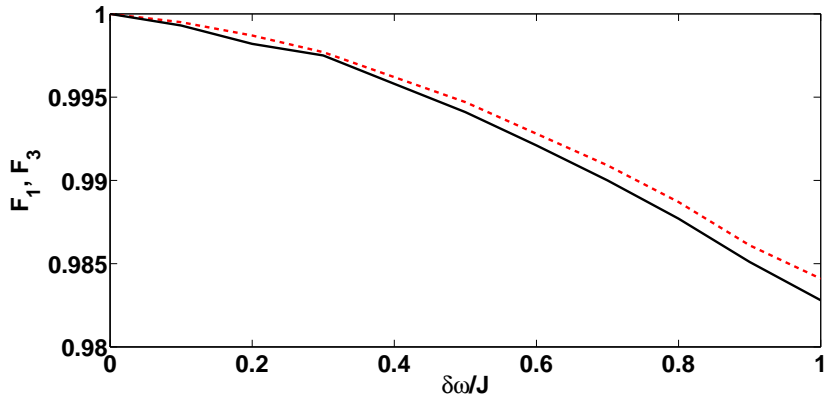}
\caption{Fidelities $F_1$ (continuous) and $F_3$ (dashed) at time $T=\pi/2J$ are shown as a function $\delta\omega/J$. We choose $(\omega/2\pi,\omega_{q3}/2\pi)=(6.92, 7)$THz, $(J/2\pi, g/2\pi)=(7,23.21)$GHz and $N=10$.}
\label{FluctuationOmegaVsFidelity}
\end{center}
\end{figure}

In a realistic situation, apart from dissipation, another factor that can reduce the fidelity of transfer  is disorder in the system. As seen in the previous section, a suitable combination of coupling strengths as is given in Eqn. \ref{EngineeredCoupling} provides perfect transfer of information if the cavities and atoms are in resonance. However, in an experimental situation, precise control of coupling strengths and resonance frequencies may not be possible. Hence, we consider small disorder in the coupling strengths as well as in the resonance frequencies. We first consider the effect of coupling disorder. Let the coupling strength $J_k$ fluctuates around $J_k\pm \frac{\delta J}{2}$ such that $\delta J$ is the width of the disorder. Here $J_k$ is the coupling strength between the $k$th and $(k+1)$th cavities.  So, $J_k$ has to be chosen randomly between $J_k-\frac{\delta J}{2}$ and $J_k+\frac{\delta J}{2}$. Also, we assume that the atom-cavity coupling strength $g$ fluctuates within a range $\delta g$. For simplicity, we take $\delta g=\delta J$. For a given set of coupling strengths that are chosen randomly from within the aforementioned interval, we calculate the fidelity of transfer at $T=\pi/2J$ and take average over all the sets of coupling strengths.  We take 1000 realizations, \textit{i.e.}, 1000 sets of coupling strengths are chosen randomly from  within the interval. Fig. \ref{FluctuationVsFidelity} shows average fidelities $F_1$ and $F_3$ as a function of the percentage of the disorder. Percentage of the disorder is defined as $\frac{\delta J}{\langle J\rangle}\times 100$, where $\langle J\rangle =(\sum_{k=1}^{N-1} J_k )/(N-1)$ is the average of all the coupling strengths. We set $J/2\pi=7$ GHz and the width of the disorder $\delta J$ is varied up to $30 \%$ of $\langle J \rangle$. As can be seen in figure, the fidelity is more than 0.95 even in the presence of large disorder in coupling strengths.  As, $F_1=F_2$ and $F_3=F_4$, we only plot $F_1$ and $F_3$ in the figure.

Now, we consider the disorder in the resonance frequencies. Let the cavity resonance frequencies are fluctuating around $\omega\pm \delta\omega/2$. Hence, the width of the disorder is $\delta\omega$. We plot the average fidelities $F_1$ and $F_3$ in Fig. \ref{FluctuationOmegaVsFidelity} as a function of $\delta\omega/J$. The average is taken over 1000 realizations, \textit{i.e.,} the set of resonance frequencies are chosen 1000 times from within the aforementioned interval. As can be seen in the figure, the fidelity of information transfer is high even in the presence of large detuning between the cavities. Hence, the scheme is robust in the presence of experimental imperfections like coupling disorder and frequency disorder.
\section{Summary}\label{Summary}    
Quantum dense coding is one of the most important protocols for realizing quantum communication, where two bits of classical information  can be transferred by sending only one quantum bit. We have proposed an experimentally realizable quantum dense coding protocol \textit{via} entanglement swapping in a cavity array containing certain number of atoms. By suitably choosing the inter-cavity coupling strengths and the atom-cavity coupling strengths, the fidelity for transfer of the information from the sender to receiver is high even in the presence of dissipation and disorder. It is possible to transfer the information with fidelity more than $0.95$ in the context of photonic crystal cavities and superconducting resonators. Fidelities can be improved further by improving the quality factor of the cavities and increasing the coupling strengths. Our scheme is also robust against experimental imperfections such as disorder in the coupling strengths and resonance frequencies. Recent progress in the fabrication of high quality cavities  suggests that our scheme may be a good candidate for realizing quantum dense coding protocol in the context of cavity QED. The proposed scheme can also be implemented in the context of spin chains, Josephson junction array, quantum dot array, optomechanical cavity array, etc. We believe that our scheme will be of use for realizing future quantum communication.     
\section*{Acknowledgement}
NM acknowledges Indian Institute of Technology Kanpur for postdoctoral fellowship.

\appendix
\numberwithin{equation}{section}
\makeatletter 
\newcommand{\section@cntformat}{Appendix \thesection:\ }
\makeatother
\setcounter{equation}{0}
\renewcommand{\theequation}{A{\arabic{equation}}}
\section*{Appendix A: Time evolution of the state} \label{timeevolution}
The Hamiltonian given in Eqn. \ref{Hamiltonian} in interaction picture is
\begin{align}\label{InteractionHamiltonian}
H_{\text{int}} &=\sum_{k=1}^{N-1} \sqrt{(k+1)(N+1-k)}J(a_k^\dagger a_{k+1}+a_k a_{k+1}^\dagger) \nonumber\\
+&\sqrt{N+1}J [a_1^\dagger (\sigma_-)_{q1}+a_1 (\sigma_+)_{q1}+a_N^\dagger (\sigma_-)_{q2}+a_N (\sigma_+)_{q2}].
\end{align}
The matrix form of the interaction Hamiltonian in the basis $\{ \ket{e}_1\ket{\text{vac}}\ket{g}_2\}$,$\{ \ket{g}_1\ket{1}\rangle\ket{g}_2\}$, $\{ \ket{g}_1\ket{2}\rangle\ket{g}_2\}$,......, $\{ \ket{g}_1\ket{N}\rangle\ket{g}_2\}$, $\{ \ket{g}_1\ket{\text{vac}}\ket{e}_2\}$  is
\begin{align}
H_{\text{int}}=J\left[
\begin{array}{ccccc}
0 & \sqrt{N+1} & 0 & \cdots & 0\\
\sqrt{N+1} & 0 & \sqrt{2N} & \cdots & 0\\
0  &  \sqrt{2N} & 0 & \cdots & 0\\
\vdots & \vdots & \vdots & \cdots  & \sqrt{N+1}\\
0 & 0 & 0  &  \sqrt{N+1} & 0
\end{array}\right].
\end{align}
One can see that this Hamiltonian is equivalent to the Hamiltonian corresponding to two coupled cavities having $(N+1)$ quanta. The Hamiltonian for the two coupled cavities is $\tilde{H}_{\text{int}}=J(\tilde{a}_1^\dagger \tilde{a}_2+\tilde{a}_1 \tilde{a}^\dagger_2)$. Also, the basis states of both the systems can be mapped to each other, \textit{i.e.}, $\{ \ket{e}_1\ket{\text{vac}}\ket{g}_2\}\rightarrow \ket{N+1,0}$, $\{ \ket{g}_1\ket{1}\rangle\ket{g}_2\}\rightarrow \ket{N,1}$, $\{ \ket{g}_1\ket{2}\rangle\ket{g}_2\} \rightarrow \ket{N-1,2}$,......, $\{ \ket{g}_1\ket{N}\rangle\ket{g}_2\}\rightarrow \ket{1,N}$ and $\{ \ket{g}_1\ket{\text{vac}}\ket{e}_2\} \rightarrow \ket{0,N+1}$. Similar kind duality has already been used in the context of cavity array \cite{Meher}.\\

As the Hamiltonians are equivalent, the time evolution of the states is also equivalent. Evolution of the state $\ket{N+1-n,n}$ under the interaction Hamiltonian $\tilde{H}_{int}$ is
\begin{align}\label{A3}
e^{-i\tilde{H}_\text{int}t}\ket{N+1-n,n}=&e^{-i\tilde{H}_\text{int}t} \frac{\tilde{a}_1^{\dagger N+1-n} \tilde{a}_2^{\dagger n}}{\sqrt{(N+1-n)! n!}}\ket{0,0}.
\end{align}
Using Baker-Hausdorf lemma \cite{Gerry}, we obtained
$e^{-i\tilde{H}_\text{int}t} \tilde{a}_1 e^{-i\tilde{H}_\text{int}t}=\tilde{a}_1\cos Jt+i\tilde{a}_2\sin Jt$,
$e^{-i\tilde{H}_\text{int}t} \tilde{a}_2 e^{-i\tilde{H}_\text{int}t}=\tilde{a}_2\cos Jt+i\tilde{a}_1\sin Jt$. Using these relations, Eqn. \ref{A3} becomes
\begin{align}
&e^{-i\tilde{H}_\text{int}t}\ket{N+1-n,n}=\nonumber\\
& \frac{(\tilde{a}_1^\dagger\cos Jt-i\tilde{a}_2^\dagger\sin Jt)^{N+1-n} (\tilde{a}_2^\dagger\cos Jt-i\tilde{a}_1^\dagger\sin Jt)^{ n}}{\sqrt{(N+1-n)! n!}} \ket{0,0}.
\end{align}
Using binomial expansion, we arrived at
\begin{align*}
&e^{-i\tilde{H}_\text{int}t}\ket{N+1-n,n}=\nonumber\\
&\sum_{k=0}^{N+1-n}\sum_{l=0}^{n}~^{N+1-n}C_k ~^nC_l (\cos Jt)^{N+1-(k+l)}(-i\sin Jt)^{k+l}\\
&\frac{\sqrt{(N+1-(n+k-l))!(n+k-l)!}}{\sqrt{(N+1-n)!n!}}\ket{N+1-(n+k-l),n+k-l}.
\end{align*}
For $n=0$, the evolved state becomes
\begin{align*}
e^{-i\tilde{H}_\text{int}t}\ket{N+1,0}=&\sum_{k=0}^{N+1} C_k (\cos Jt)^{N+1-k}(-i\sin Jt)^{k}\ket{N+1-k,k},
\end{align*}
where the expression for $C_k$ is given in Eqn. \ref{Ck}.
Equivalently, in the cavity array, the evolved state becomes
\begin{align}
&e^{-iH_{\text{int}}t}\ket{e}_1\ket{\text{vac}}\ket{g}_2= \left[(\cos Jt)^{N+1}\ket{e}_1\ket{\text{vac}}\ket{g}_2\right.\nonumber\\ 
&\left. + \sum_{k=1}^{N} C_k\ket{g}_1\ket{k}\rangle\ket{g}_2 \right. \left.+(-i\sin Jt)^{N+1}\ket{g}_1\ket{\text{vac}}\ket{e}_2\right].
\end{align}

\section*{References}


\begin{thebibliography}{10}
\expandafter\ifx\csname url\endcsname\relax
  \def\url#1{{\tt #1}}\fi
\expandafter\ifx\csname urlprefix\endcsname\relax\def\urlprefix{URL }\fi
\providecommand{\eprint}[2][]{\url{#2}}

\bibitem{Bennett}
Bennett C~H and Wiesner S~J 1992 {\em Phys. Rev. Lett.\/} {\bf 69}(20)
  2881--2884
  \urlprefix\url{https://link.aps.org/doi/10.1103/PhysRevLett.69.2881}

\bibitem{Mattle}
Mattle K, Weinfurter H, Kwiat P~G and Zeilinger A 1996 {\em Phys. Rev. Lett.\/}
  {\bf 76}(25) 4656--4659
  \urlprefix\url{https://link.aps.org/doi/10.1103/PhysRevLett.76.4656}

\bibitem{Ban}
Ban M 1999 {\em Journal of Optics B: Quantum and Semiclassical Optics\/} {\bf
  1} L9--L11

\bibitem{Liang}
Qiu L, Wang A~M and Ma X~S 2007 {\em Physica A: Statistical Mechanics and its
  Applications\/} {\bf 383} 325 -- 330 ISSN 0378-4371

\bibitem{Simayi}
Simayi S, Abulizi A, Yaermaimaiti M, Cai J~T and Qiao P~P 2011 {\em Chinese
  Physics B\/} {\bf 20} 050305

\bibitem{Xu}
Xu H~Y and Yang G~H 2017 {\em International Journal of Theoretical Physics\/}
  {\bf 56} 2803--2810 ISSN 1572-9575
  \urlprefix\url{https://doi.org/10.1007/s10773-017-3445-0}

\bibitem{Huang}
Huang H 2009 {\em International Journal of Theoretical Physics\/} {\bf 48} 3491
  ISSN 1572-9575 \urlprefix\url{https://doi.org/10.1007/s10773-009-0153-4}

\bibitem{Xiu}
Lin X~M, Zhou Z~W, Xue P, Gu Y~J and Guo G~C 2003 {\em Physics Letters A\/}
  {\bf 313} 351 -- 355 ISSN 0375-9601

\bibitem{Ye}
Ye L and Guo G~C 2005 {\em Phys. Rev. A\/} {\bf 71}(3) 034304
  \urlprefix\url{https://link.aps.org/doi/10.1103/PhysRevA.71.034304}

\bibitem{Xue}
Xue Z~Y, min Yi Y and liang Cao Z 2006 {\em Journal of Modern Optics\/} {\bf
  53} 2725--2732

\bibitem{Jia}
Chun-Xia J and Zhao-Hui P 2008 {\em Communications in Theoretical Physics\/}
  {\bf 50} 1113--1116

\bibitem{Juan}
Juan H, Liu Y and Zhi-Xiang N 2008 {\em Chinese Physics B\/} {\bf 17}
  1597--1600

\bibitem{Nie}
Nie Y~y, Li Y~h, Wang X~p and Sang M~h 2013 {\em Quantum Information
  Processing\/} {\bf 12} 1851--1857 ISSN 1573-1332
  \urlprefix\url{https://doi.org/10.1007/s11128-012-0499-z}

\bibitem{Reiserer}
Reiserer A and Rempe G 2015 {\em Rev. Mod. Phys.\/} {\bf 87}(4) 1379--1418

\bibitem{Wu}
Wu Q and Yang M 2008 {\em International Journal of Theoretical Physics\/} {\bf
  47} 3139--3143 ISSN 1572-9575

\bibitem{Meher}
Meher N, Sivakumar S and Panigrahi P~K 2017 {\em Scientific Reports\/} {\bf 7}
  9251-- ISSN 2045-2322

\bibitem{Almeida}
Almeida G~M~A, Ciccarello F, Apollaro T~J~G and Souza A~M~C 2016 {\em Phys.
  Rev. A\/} {\bf 93}(3) 032310

\bibitem{YangLiu}
Liu Y and Zhou D~L 2015 {\em New Journal of Physics\/} {\bf 17} 013032

\bibitem{Yung}
Yung M~H and Bose S 2005 {\em Phys. Rev. A\/} {\bf 71}(3) 032310

\bibitem{Meher_2019}
Meher N 2019 {\em Journal of Physics B: Atomic, Molecular and Optical
  Physics\/} {\bf 52} 205502

\bibitem{Neto}
de~Moraes~Neto G~D, Andrade F~M, Montenegro V and Bose S 2016 {\em Phys. Rev.
  A\/} {\bf 93}(6) 062339

\bibitem{Hua}
Hua M, Tao M~J and Deng F~G 2016 {\em Scientific Reports\/} {\bf 6} 22037--

\bibitem{Meher2019}
Meher N 2020 {\em International Journal of Theoretical Physics\/} {\bf 59}
  218--228 ISSN 1572-9575
  \urlprefix\url{https://doi.org/10.1007/s10773-019-04314-1}

\bibitem{Leons}
Leonski W and Miranowicz A 2004 {\em Journal of Optics B: Quantum and
  Semiclassical Optics\/} {\bf 6} S37

\bibitem{Liew}
Liew T~C~H and Savona V 2013 {\em New Journal of Physics\/} {\bf 15} 025015

\bibitem{Browne}
Browne D~E and Plenio M~B 2003 {\em Phys. Rev. A\/} {\bf 67}(1) 012325

\bibitem{Miry}
Miry S~R, Tavassoly M~K and Roknizadeh R 2015 {\em Quantum Information
  Processing\/} {\bf 14} 593--606 ISSN 1573-1332

\bibitem{Yurk}
Yurke B and Stoler D 1986 {\em Phys. Rev. Lett.\/} {\bf 57}(1) 13--16

\bibitem{Rojan}
Rojan K, Reich D~M, Dotsenko I, Raimond J~M, Koch C~P and Morigi G 2014 {\em
  Phys. Rev. A\/} {\bf 90}(2) 023824

\bibitem{WeiWei}
Wei W and Guang-can G 1998 {\em Acta Physica Sinica (Overseas Edition)\/} {\bf
  7} 174

\bibitem{Yanhua}
Wang Y, Wan J, Zou B, Zhang J and Zhu Y 2013 {\em Journal of Physics:
  Conference Series\/} {\bf 414} 012001

\bibitem{Meher3}
Meher N and Sivakumar S 2018 {\em Quantum Information Processing\/} {\bf 17}
  233 ISSN 1573-1332

\bibitem{Meher2}
Meher N and Sivakumar S 2016 {\em J. Opt. Soc. Am. B\/} {\bf 33} 1233--1241

\bibitem{Schmidt}
Schmidt S, Gerace D, Houck A~A, Blatter G and T\"ureci H~E 2010 {\em Phys. Rev.
  B\/} {\bf 82}(10) 100507

\bibitem{Asad}
Asadian A, Manzano D, Tiersch M and Briegel H~J 2013 {\em Phys. Rev. E\/} {\bf
  87}(1) 012109

\bibitem{Meher_2020}
Meher N and Sivakumar S 2020 {\em J. Opt. Soc. Am. B\/} {\bf 37} 138--147
  \urlprefix\url{http://josab.osa.org/abstract.cfm?URI=josab-37-1-138}

\bibitem{Xuereb}
Xuereb A, Imparato A and Dantan A 2015 {\em New Journal of Physics\/} {\bf 17}
  055013

\bibitem{Imamoglu}
Imamoglu A, Schmidt H, Woods G and Deutsch M 1997 {\em Phys. Rev. Lett.\/} {\bf
  79}(8) 1467--1470

\bibitem{Birnbaum}
Birnbaum K~M, Boca A, Miller R, Boozer A~D, Northup T~E and Kimble H~J 2005
  {\em Nature\/} {\bf 436} 87--

\bibitem{Tang}
Tang J, Geng W and Xu X 2015 {\em Scientific Reports\/} {\bf 5} 9252--

\bibitem{Shen}
Shen H~Z, Zhou Y~H and Yi X~X 2015 {\em Phys. Rev. A\/} {\bf 91}(6) 063808

\bibitem{Miranowicz}
Miranowicz A, Bajer J~c~v, Paprzycka M, Liu Y~x, Zagoskin A~M and Nori F 2014
  {\em Phys. Rev. A\/} {\bf 90}(3) 033831

\bibitem{Zhou}
Zhou L, Gong Z~R, Liu Y~x, Sun C~P and Nori F 2008 {\em Phys. Rev. Lett.\/}
  {\bf 101}(10) 100501

\bibitem{Qin}
Qin W and Nori F 2016 {\em Phys. Rev. A\/} {\bf 93}(3) 032337

\bibitem{Yang}
Yang M, Zhao Y, Song W and Cao Z~L 2005 {\em Phys. Rev. A\/} {\bf 71}(4) 044302
  \urlprefix\url{https://link.aps.org/doi/10.1103/PhysRevA.71.044302}

\bibitem{Ying}
Ying-Qiao Z, Xing-Ri J and Shou Z 2005 {\em Chinese Physics\/} {\bf 14}
  1732--1735

\bibitem{Xiu2007}
Xiu L, Hong-Cai L, Xiu-Min L, Xing-Hua L and Rong-Can Y 2007 {\em Chinese
  Physics\/} {\bf 16} 1209--1214

\bibitem{Yogesh}
Joglekar Y~N, Thompson C, Scott D~D and Vemuri G 2013 {\em The European
  Physical Journal Applied Physics\/} {\bf 63}

\bibitem{Divincenzo2}
DiVincenzo D~P 1995 {\em Phys. Rev. A\/} {\bf 51}(2) 1015--1022

\bibitem{Divincenzo3}
DiVincenzo D~P 2000 {\em Fortschritte der Physik\/} {\bf 48} 771--783

\bibitem{Barenco2}
Barenco A, Deutsch D, Ekert A and Jozsa R 1995 {\em Phys. Rev. Lett.\/} {\bf
  74}(20) 4083--4086
  \urlprefix\url{https://link.aps.org/doi/10.1103/PhysRevLett.74.4083}

\bibitem{DiVincenzo}
DiVincenzo D~P 1995 {\em Science\/} {\bf 270} 255--261 ISSN 0036-8075
  \urlprefix\url{https://science.sciencemag.org/content/270/5234/255}

\bibitem{Carmichael}
Carmichael H~J 1999 {\em {Statistical Methods in Quantum Optics 1: Master
  Equations and Fokker-Planck Equations}\/} (Springer)

\bibitem{Lindblad}
Lindblad G 1976 {\em Commun.Math. Phys.\/} {\bf 48}(2) 195452

\bibitem{Noto}
Notomi M, Tanabe T, Shinya A, Kuramochi E, Taniyama H, Mitsugi S and Morita M
  2007 {\em Opt. Express\/} {\bf 15} 17458--17481

\bibitem{Noto2}
Notomi M, Kuramochi E and Tanabe T 2008 {\em Nature Photonics\/} {\bf 2} 741--

\bibitem{Vasco}
Vasco J~P and Savona V 2018 {\em New Journal of Physics\/} {\bf 20} 075002

\bibitem{Arka}
Majumdar A, Rundquist A, Bajcsy M, Dasika V~D, Bank S~R and Vu\ifmmode
  \check{c}\else \v{c}\fi{}kovi\ifmmode~\acute{c}\else \'{c}\fi{} J 2012 {\em
  Phys. Rev. B\/} {\bf 86}(19) 195312

\bibitem{Kuhr}
Kuhr S, Gleyzes S, Guerlin C, Bernu J, Hoff U~B, Deleglise S, Osnaghi S, Brune
  M, Raimond J~M, Haroche S, Jacques E, Bosland P and Visentin B 2007 {\em
  Applied Physics Letters\/} {\bf 90} 164101

\bibitem{Arka2}
Majumdar A, Rundquist A, Bajcsy M and Vu\ifmmode \check{c}\else
  \v{c}\fi{}kovi\ifmmode~\acute{c}\else \'{c}\fi{} J 2012 {\em Phys. Rev. B\/}
  {\bf 86}(4) 045315

\bibitem{Xiao}
Xiao Y~F, Gao J, Zou X~B, McMillan J~F, Yang X, Chen Y~L, Han Z~F, Guo G~C and
  Wong C~W 2008 {\em New Journal of Physics\/} {\bf 10} 123013

\bibitem{Tiecke}
Tiecke T~G, Thompson J~D, de~Leon N~P, Liu L~R, Vuletic V and Lukin M~D 2014
  {\em Nature\/} {\bf 508} 241
  \urlprefix\url{https://doi.org/10.1038/nature13188}

\bibitem{Du}
Du H, Zhang X, Chen G, Deng J, Chau F~S and Zhou G 2016 {\em Scientific
  Reports\/} {\bf 6} 24766

\bibitem{Ma}
Ma S~l, Xie J~k, Li X~k and Li F~l 2019 {\em Phys. Rev. A\/} {\bf 99}(4) 042317

\bibitem{Chen}
Chen Q~M, Liu Y~x, Sun L and Wu R~B 2018 {\em Phys. Rev. A\/} {\bf 98}(4)
  042328

\bibitem{Li2}
Li H, Wang Y, Wei L, Zhou P, Wei Q, Cao C, Fang Y, Yu Y and Wu P 2013 {\em
  Chinese Science Bulletin\/} {\bf 58} 2413--2417

\bibitem{Gerry}
Gerry C and Knight P 2004 {\em {Introductory Quantum Optics}\/} (Cambridge
  University Press) ISBN 052152735X

\end{thebibliography}
\providecommand{\newblock}{}

\end{document}